# A Delay-Based Approach for Congestion Avoidance in Interconnected Heterogeneous Computer Networks

Raj Jain
Digital Equipment Corporation
550 King St. (LKG1-2/A19)
Littleton, MA 01460

## Abstract

In heterogeneous networks, achieving congestion avoidance is difficult because the congestion feedback from one subnetwork may have no meaning to sources on other subnetworks. We propose using changes in round-trip delay as an implicit feedback. Using a black-box model of the network, we derive an expression for the optimal window as a function of the gradient of the delay-window curve.

The problems of selfish optimum and social optimum are also addressed. It is shown that without a careful design, it is possible to get into a race condition during heavy congestion, where each user wants more resources than others, thereby leading to a diverging condition.

It is shown that congestion avoidance using round-trip delay is a promising approach. The approach has the advantage that there is absolutely no overhead for the network itself. It is exemplified by a simple scheme. The performance of the scheme is analyzed using a simulation model. The scheme is shown to be efficient, fair, convergent, and adaptive to changes in network configuration.

The scheme as described works only for networks which can be modeled with queueing servers with constant service times. Further research is required to extend it for implementation in practical networks. Several directions for future research have been suggested.

## 1 Introduction

Most networking architectures have schemes for congestion control. Digital's Networking Architecture (DNA) [4] uses a timeout-based congestion control [14] and square root input buffer limiting [7]. IBM's System Networking Architecture (SNA) uses congestion bits called *change window indicator* (CW) and *reset window indicator* (RW) in packets flowing in the reverse direction to ask sources to reduce the load during congestion [1]. DARPA's TCP/IP networks use source quench messages in a similar manner. In general, all congestion schemes consist of a feedback signal from the network to the users (timeout, bits, or messages) and a load-control mechanism exercised by the users (reduced window or rate).

Today, we have several leading networking architectures, each with its own philosophy, assumptions, and objectives. A communications medium, by definition, cannot stay aloof for long. As networking becomes popular, we want to communicate farther and farther and by necessity need to use intermediate networks that may or may not have been designed with the same philosophy.

In a network consisting of heterogeneous subnetworks, the congestion feedback from one subnetwork may have no meaning to sources on other subnetworks. The problem is similar to that of deciphering traffic control signs in a foreign country. Finding an effective means of feedback in such networks is not trivial. The controlling mechanism in such networks have to rely on implicit feedback mechanism such as timeouts, which happen during congestion in all architectures.

We are concerned here with *congestion avoidance* rather than *congestion control* in heterogeneous networks. Briefly, a congestion avoidance scheme allows a network to operate in the region of low delay and high throughput [16]. We will elaborate on this point in the next section. The approach that we propose here is called 'Congestion Avoidance using Round-trip Delay' or **CARD**. The approach is based on the simple fact that as the load on the network increases and queues build up, the round-trip delay increases. Most transport protocols measure round-trip delays to set timers for timeout [11] and can use this information to adjust their load on the network.

The delay-based scheme proposed in this paper is not intended to replace the bit-based binary feedback scheme, we proposed earlier in [15,19,20]. The bit-based scheme is a fully worked out scheme and has been tested via simulations to perform well under a wide variety of circumstances. The delay-based scheme proposed here is only an example of an approach, which, we feel, is a promising direction for researchers to explore. The results presented here represent only our initial effort in this direction. Further work is required to design a practical delay-based scheme that can be implemented in real networks.

## 2 Congestion Avoidance

Figure 1 shows general patterns of response time and throughput of a network as a function of its load. If the load is small, throughput generally keeps up with the load. As the load increases, throughput increases. After the load reaches the network capacity, throughput stops increasing. If the load is increased any further, the queues start building, potentially resulting in packets being dropped. Throughput may suddenly drop when the load increases beyond this point and the network is said to be congested. The delay (or response-time) curve follows a similar pattern. At first the response time increases little with the load. When the queues start building up, the response time increases linearly until finally, as the queues start overflowing, the response time increases drastically.

The point at which throughput approaches zero is called the **cliff** due to the fact that throughput falls off rapidly after this point. We use the term **knee** to describe the point after which the increase in the throughput is small, but the increase in response time is significant.

A scheme that allows the network to operate at the knee is called a **congestion avoidance** scheme, as distinguished from a *congestion control* scheme that tries to keep the network operating in the zone to the left of the cliff.

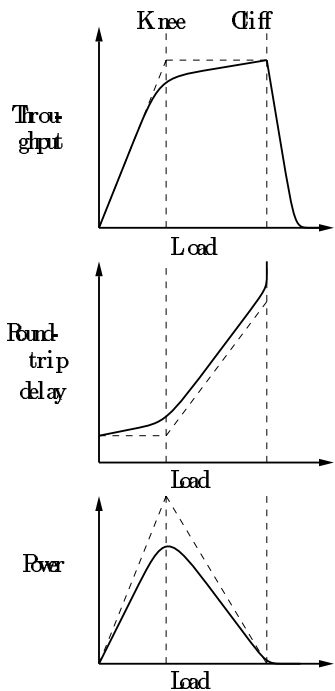

Figure 1: Network performance as a function of the load. Broken curves indicate performance with deterministic service and interarrival times.

The number of packets in a path, when it is operating at the knee, is called **knee capacity** or the **pipe size** of the path. We elaborate further on these concepts in [15,16].

## 3 Black-box Approach

The delay-based approach proposed in this paper is, what we call, a *black-box* approach. It treats the network as a black box, which does not give any explicit feedback. We need to deduce the network load based solely on the information available outside the network. Examples of such information are timeouts, decreased throughput, or increased delay. Black-box congestion control schemes using timeouts are already being used in several architectures including DNA, OSI/TP4 [14], and TCP [2].

Black-box schemes have no *explicit* feedback and are therefore also called *implicit* feedback schemes. Such schemes may be used even if a network already has an explicit feedback scheme. The latter works only for those resources that can send the feedback. O-ten it happens that even though the network does have an explicit feedback scheme, some congested resources cannot send such a feedback. For example, LAN bridges operate transparently at the data-link layer and cannot set congestion bits which are at the network layer. A bridge, if congested, can only drop packets without notifying the source. A similar argument applies if other data-link level elements, such as LAN adapters, are congested, but the the feedback is implemented at a higher layer.

The advantages of black-box schemes for heterogeneous networks are obvious. Since there are no universally agreed explicit feedback signals, one subnetwork may not know about the feedback signals from other subnetworks.

Black-box schemes are not an alternative to explicit feedback schemes. They are complementary. With proper information, any system can be made to perform better than without any information. Implicit feedback schemes increase the amount of information available by adding implicit feedback to the explicit feedback, if available.

Black-box schemes are *zero network overhead* schemes. The flow control, congestion control, and congestion avoidance mechanisms, while essential for network operation, are actually overheads since they themselves consume the very resource they are suppose to allocate. It is possible to get into a 'thrashing' situation in which all resources are totally consumed by the control messages with nothing left for the users. The network architects are therefore constantly looking for ways to minimize these overheads [12]. Xon/Xoff flow control and timeout-based congestion control are examples of ways to achieve flow and congestion control with minimal or no explicit feedback. In this paper, we report preliminary results of our efforts to design a mechanism for congestion avoidance that requires no explicit feedback from the network.

## 4 Optimal Window Size

Figure 2 shows the black-box view of a network of several LANs, terrestrial and satellite links. Users are not aware of the internals of the network. They treat it as a black-box. As they increase their load on the network, the delay increases and based on this delay their task is to determine the optimal load.

The end-to-end delay experienced by packets trans-

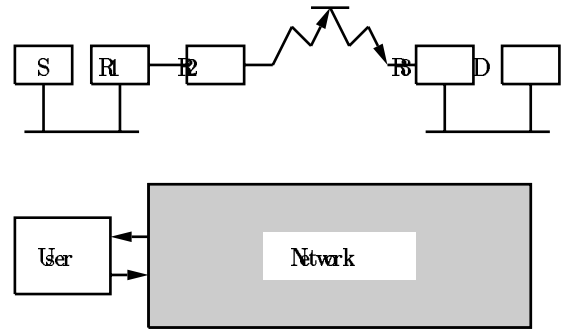

Figure 2: A black-box view of the network.

mitted by an end system is a function of several parameters including the following:

1. Window size (or load) of the end system
2. Packet or train interarrival pattern [13]
3. Number of network resources used
4. Service time distribution of individual resources
5. Number of other end systems sharing the resources
6. Window size and interarrival pattern of other end systems.

The problem of interpreting the 'delay signals' is quite complex unless we make some simplifying assumptions. Let us first assume that there are no other users on the network. This eliminates the fairness considerations and simplifies the efficiency considerations. Also, we assume that the source uses a window flow control mechanism. Treating the network as a black-box, the source can measure the network delay and throughput for any given window. It can also compute the 'power,' which is defined as the ratio of throughput and delay [5,17]. By plotting the power as a function of window size, it can determine the window at which the power is maximum. This is the knee.

The procedure as outlined above can be further simplified in several different ways. To explain these alternatives we need to define a number of symbols and explain the notation. The following symbols will be used:

$$
\begin{aligned}
W &= \text{Window} = \text{Number of packets in the network} \\
T &= \text{Throughput in packets per unit time} \\
D &= \text{Round-trip delay} \\
P &= \text{Power} = \frac{T^\alpha}{D} \\
\alpha &= \text{Exponent used in defining power} \\
\hat{x} &= \text{The optimal value of } x, \text{ i.e., the value of } x \text{ at the knee. Here } x=D,\ P,\ T, \text{ or } W.
\end{aligned}
$$

The round-trip delay $D$ and the throughput $T$ are both functions of the window $W$:

$$D = f_D(W)$$

$$T = f_T(W)$$

The power is defined as the ratio of throughput and delay:

$$P = \frac{T^\alpha}{D}$$

Here, $\alpha$ is a parameter chosen by system designers. Its impact will be clear shortly.

$$\log(P) = \alpha \log(T) - \log(D)$$

At the point of maximum power, i.e., at the knee:

$$\frac{dP}{P} = \alpha \frac{dT}{T} - \frac{dD}{D} = 0$$

or,

$$\alpha \frac{dT}{T} = \frac{dD}{D}$$

Thus, at the knee, the relative (percentage) increase in delay is $\alpha$ times the relative increase in throughput. If we choose $\alpha = 1$, the percentage increase in delay is equal to the percentage increase in throughput at the knee. Before the knee:

$$\frac{dD}{D} < \frac{dT}{T}$$

the relative increase in delay is smaller than the relative gain in throughput. After the knee:

$$\frac{dD}{D} > \frac{dT}{T}$$

the relative increase in delay is larger than the relative gain in throughput.

If we want to allow higher relative increase in *delay* at the knee, we can choose $\alpha > 1$. Similarly, $\alpha < 1$ can be used to achieve higher relative increase in *throughput* at the knee.

For window flow controlled networks, the user's throughput $T$ is $W$ packets per round-trip delay, or

$$T = \frac{W}{D}$$

and therefore,

$$\log(P) = \alpha \log(W) - (1+\alpha)\log(D)$$

$$\frac{dP}{P} = \alpha \frac{dW}{W} - (1+\alpha)\frac{dD}{D} = 0$$

By solving the above condition for $W$, we get the optimal window size $\hat{W}$ as:

$$\hat{W} = \frac{\alpha}{1+\alpha}\left(\frac{D}{\frac{dD}{dW}}\right) \quad (1)$$

Since all of the quantities on the right hand side of the above equation are known, we can compute the optimal window size $\hat{W}$.

The results so far are valid for all networks or resources since we have made no assumptions about the behavior of the internal components of the network, deterministic or probabilistic distributions of service times, or linear or nonlinear behavior of the delay versus window curve.

If there are no other users on the network, it provides a way for one user to determine the knee using the measured delay and the gradient $dD/dW$ of the delay-window curve. This is the key formula leading us to hope that a black-box approach to congestion avoidance may be feasible.

The value of $\hat{W}$ as computed using equation (1) gives the *optimal direction* for window adjustment. If the current window $W$ is less than $\hat{W}$, then we should increase the window. Similarly, if the current window $W$ is greater than $\hat{W}$, we should decrease the window. The exact difference between $W$ and $\hat{W}$ may or may not be meaningful. For example, if the gradient $dD/dW$ is zero at a particular $W$, $\hat{W}$ is infinite indicating that $W$ should be increased. This should *not* be interpreted to mean that the path has an infinite knee capacity. At different values of window $W$, the computed $\hat{W}$ may be different, but in each case, it points in the right direction. In short, only the sign, and not the magnitude, of the difference ($\hat{W} - W$), is meaningful.

One possible way to determine the correct direction of window adjustment is to use the **normalized delay gradient (NDG)** which, we define, as the ratio:

$$\text{Normalized delay gradient} = \frac{dD/dW}{D/W}$$

If the load is low, NDG is low. If the load is high, NDG is high. At the knee, NDG is one-half as can be seen by using equation 1:

$$\frac{dD/dW}{D/W} = \frac{\alpha}{1+\alpha}$$
$$= \frac{1}{2} \quad \text{if } \alpha = 1$$

Thus, by computing NDG, we may be able to decide whether to increase or decrease the window.

### 4.1 Selfish Optimum versus Global Optimum

For multiuser cases, the application of equation (1) is not as straightforward as it may appear. In particular, there are two different optimal operating points: social and selfish.

Given $n$ users sharing a single path, the system throughput is a function of the sum of the windows of all $n$ users:

$$T = \frac{\sum_{i=1}^{n} W_i}{D}$$

Here, $W_i$ is the window of the $i^{th}$ user, and $D$ is the common delay experienced by each of the $n$ users. The system power is defined on the basis of system throughput:

$$P = \frac{T^\alpha}{D} = \frac{(\sum_{i=1}^{n} W_i)^\alpha}{D^{1+\alpha}} = D^{-1-\alpha} \left(\sum_{i=1}^{n} W_i\right)^\alpha$$

The point of maximum system power is given by a set of $n$ equations like the following:

$$\frac{\partial P}{\partial W_i} = -(1+\alpha) D^{-2-\alpha} \frac{\partial D}{\partial W_i} \left(\sum_{i=1}^{n} W_i\right)^\alpha$$
$$+ D^{-1-\alpha} \alpha \left(\sum_{i=1}^{n} W_i\right)^{\alpha-1}$$
$$= 0$$

or,

$$\sum_{i=1}^{n} W_i = \frac{\alpha}{1+\alpha} \left(\frac{D}{\frac{\partial D}{\partial W_i}}\right)$$

or,

$$\hat{W}_i = \frac{\alpha}{1+\alpha} \left(\frac{D}{\frac{\partial D}{\partial W_i}}\right) - \sum_{j \neq i}^{n} \hat{W}_j \quad (2)$$

The optimal operating point so obtained is called the **social optimum**.

Each individual user's power $P_i$ is based on the user's throughput $T_i$ and is given by

$$T_i = \frac{W_i}{D}$$

and

$$P_i = \frac{T_i^\alpha}{D} = \frac{W_i^\alpha}{D^{1+\alpha}} = D^{-1-\alpha} W_i^\alpha$$

The user's power is maximum when:

$$\frac{\partial P_i}{\partial W_i} = -(1+\alpha) D^{-2-\alpha} \frac{\partial D}{\partial W_i} W_i^\alpha + D^{-1-\alpha} \alpha W_i^{\alpha-1} = 0$$

or,

$$\hat{W}_i = \frac{\alpha}{1+\alpha} \left(\frac{D}{\frac{\partial D}{\partial W_i}}\right) \quad (3)$$

The operating point so obtained is called the **selfish optimum**. It is clear by examining equations (2) and (3) that the $\hat{W}_i$ obtained by selfish optimum is not the same as that obtained by social optimum. They may not point a user in the same direction. The two values are equal if $\sum_j \hat{W}_j = 0$, that is, if there is only one user on the network. For such a case, we can use either equation to determine the direction of window adjustment.

Social considerations would lead conscientious users to use lower windows as other users increase their windows. While selfish considerations would lead the users to use higher windows as other users increase their windows. Interestingly, this behavior is not only mathematically true as we showed above but also 'psychologically' true. People start hoarding a resource and increase their apparent demand for it if the resource becomes in short supply.

In congestion avoidance we are really interested in attaining social optimum. Selfish optimum leads to a race condition in which each user tries to maximize its power at the cost of that of the others, and the windows keep increasing without bound. Later, we will show the simulation result of one such case. Unfortunately, by examining equation (2), it is clear that to determine one's socially optimum window each user may need to know the windows of other users. A congestion avoidance policy requiring each user to inform other users of its window will cause too much overhead to be acceptable.

Fortunately, there is a special case in which knowledge of other users' windows is not required to achieve the social optimum. This case happens for deterministic networks.

## 4.2 Deterministic Networks

A deterministic computer network is one in which the packet service time at the servers is not a random variable. The service time per packet at different servers may be different but they are all fixed. Analytically, such networks can be modeled by a closed queueing network of $m$ D/D/1 servers, where $m$ is the number of queues that the packets and their acknowledgments pass through in one round trip through the network. For such networks the delay versus window curve consists of two straight line segments meeting at the knee. Before the knee, the delay is constant:

$$D(W) = \sum_{i=1}^{m} t_i$$

where, $t_i$ is the service time of the $i^{th}$ server. After the knee, the delay increases linearly:

$$D(W) = Wt_b$$

where $t_b$ is the service time of the bottleneck server, i.e.,

$$t_b = \max_i \{t_i\}$$

Fixed delay servers such as satellite links are not included in the maximum determination but are included in the summation. The two equations for delay above can be combined into one:

$$D(W) = \max \left\{ \sum_{i=1}^{m} t_i, Wt_b \right\}$$

The power is maximum at the knee, where:

$$\sum_{i=1}^{m} t_i = Wt_b$$

or,

$$W_{knee} = \frac{\sum_{i=1}^{m} t_i}{t_b} \quad (4)$$

Equation (4) for optimal window size helps us compute the knee capacity of a path:[1]

$$\text{Knee capacity of a path} \approx \frac{\text{Sum of all service times}}{\text{Bottleneck service time}}$$

For deterministic networks, $\frac{\partial D}{\partial W_i}$ and NCG are zero to the left of the knee. This property helps us achieve the social optimum in a distributed fashion. This is the basis of the congestion avoidance scheme described next.

---
[1] This expression for knee capacity is approximately valid for unbalanced probabilistic networks as well.

## 5 A Sample Scheme

The users of the network need guidelines to answer the following three questions:

1. Whether to increase or decrease the window?
2. How much should the change in window be?
3. How often to change the window?

The components of the congestion avoidance scheme which answer these questions are called decision function, increase/decrease algorithm, and decision frequency, respectively. These three components together form what is called *user policy* [16]. The delay-based schemes have no *network policy* since the network does not explicitly participate in the congestion avoidance. In the following, we describe the three components of a sample scheme in detail.

### 5.1 Decision Function

The decision function helps the user determine the direction of window adjustment. We can use NCG as the decision function. For deterministic networks, NCG is zero to the left of the knee. Given round-trip delays $D$ and $D_{old}$ at windows $W$ and $W_{old}$ respectively, the decision function consists of checking simply if NCG is zero. The exact algorithm is described below

$$\text{NCG} \leftarrow \left(\frac{D - D_{old}}{D + D_{old}}\right)\left(\frac{W + W_{old}}{W - W_{old}}\right);$$
IF (NCG $> 0$ or $W = W_{max}$)
THEN Decrease($W$)
ELSE IF (NCG $\leq 0$ or $W = W_{min}$)
THEN Increase($W$);

In the above algorithm, $W_{min}$ and $W_{max}$ are lower and upper bound on the window. The upper bound is set equal to the flow control window permitted by the receiving node based on its local buffer availability considerations. The lower bound is greater or equal to one since the window cannot be reduced to zero.

$$W_{min} \geq 1$$

$$W_{max} \geq W_{min}$$

By setting $W_{min} = W_{max}$, we can disable the window adjustment.

Note that the window must either increase or decrease at every decision point. It cannot remain constant (except when the scheme has been disabled by setting $W_{min} = W_{max}$). This is necessary since the network load is constantly changing. It is important to ensure that changes in gradient, if any, are detected as soon as possible.

Also note that instead of checking whether the change in delay $D - D_{old}$ is zero, we check whether NG is zero. The two conditions may be equivalent but we prefer the latter since NG is a dimensionless quantity and its value remains the same regardless of whether we measure delays in picoseconds or years! The difference in delay can be made to look arbitrarily small (or large) by appropriate manipulation of its units. NG is not susceptible to such manipulations.

### 5.2 Increase/Decrease Algorithm

The scheme uses additive increase and multiplicative decrease algorithms which have been shown to be the simplest alternatives leading to fairness and convergence [9, 16, 3] for multiple users starting at arbitrary window values. Thus, if the window has to be increased, we do so additively:

$$W \leftarrow W + \Delta W$$

For a decrease, window is multiplied by a factor less than one:

$$W \leftarrow cW, \quad c < 1$$

The parameters $\Delta W$ and $c$ affect the amplitude and frequency of oscillations when the system operating point approaches the knee. Recommended values of these two parameters are $\Delta W = 1$ and $c = 0.875$.

The choice of additive increase and multiplicative decrease can be briefly justified as follows. If the network is operating below the knee, all users go up equally, but, if the network is congested, the multiplicative decrease makes users with higher windows go down more than those with lower windows, making the allocation more fair. Note that $0.875 = 1 - 2^{-3}$. Thus, the multiplication can be performed without floating point hardware, and by simple logical shift instructions. The recommended values of the increase/decrease parameters lead to small oscillations and are easy to implement.

The computations should be rounded to the nearest integer. Truncation, instead of rounding, results in a slightly lower fairness.

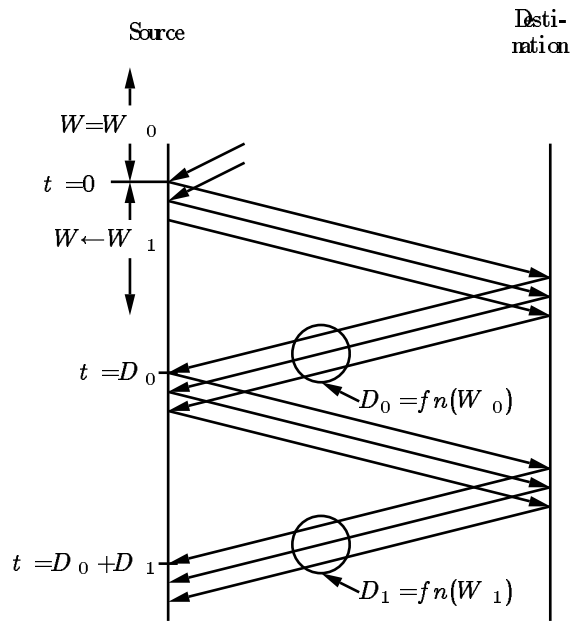

Figure 3: The round-trip delay immediately after a change of window from $W_0$ to $W_1$ corresponds to $W_0$.

### 5.3 Decision Frequency

This component helps decide how often to change the window. Changing it too often leads to unnecessary oscillations, whereas changing it infrequently leads to a system that takes too long to adapt. System control theory tells us that the optimal control frequency depends upon the feedback delay — the time between applying a control (change window) and getting feedback from the network corresponding to this control.

In computer networks, it takes one round-trip delay to affect the control, that is, for the new window to take effect and another round-trip delay to get the resulting change fed back from the network to the users. This, therefore, leads to the conclusion that windows be adjusted once every two round-trip delays (two window turns) and that only the feedback signals received in the past cycle be used in window adjustment, as shown in Figure 3.

In the procedure as outlined above, alternate delay measurements are discarded. This leads to a slight loss of information which can be avoided by a simple modification. The delay experienced by every packet is a function of the number of packets already in the network. This number is normally equal to the cur-rent window except at the point of window change. If for those packets whose sending times are recorded for round-trip delay measurements, we also record the number $W_{out}$ of packets outstanding (packets sent but not acknowledged) at the time of sending, the delay $D$ and the number $W_{out}$ have a one-to-one correspondence. Any two $\{W_{out}, D\}$ pairs can thus be used to compute NDG. This modification allows us to update window every round-trip delay. The increased information results in a faster response to the network changes. The simulation results, presented later in this paper, use this modification.

### 5.4 Initialization

The scheme does not set any requirements on the window values to be used at connection initialization. Transports can start the connections at any value and the scheme will eventually bring the load to the knee level. Later we will show simulation results to prove this. Nonetheless, starting at the minimum window value is recommended as this causes minimal affect on other users that may already be using the network.

## 6 Performance of The Scheme

We used a simulation model to study the performance of various delay-based congestion avoidance alternatives. Actually, this is the same model [10] that we had used earlier for developing the timeout-based congestion control scheme CUTE [14] and the binary feedback congestion avoidance scheme [15,19,20]. The model allows us to simulate a general computer network with several terrestrial and satellite links. Any reasonable number of users, intermediate systems, and links can be simulated. Currently the model simulates only one-way flow of packets from source to the destinations. The reverse flow of acknowledgments from the destination to source is not explicitly simulated. The source is informed instantaneously as soon as the packet is received by the destination. The model does not allow simulation of the acknowledgment withholding or path splitting. In all simulations reported here, the intermediate systems were configured with enough buffers to disable packet loss due to buffer shortage.

We simulated a number of configurations. Two of these configurations and the corresponding simulation results are described below.

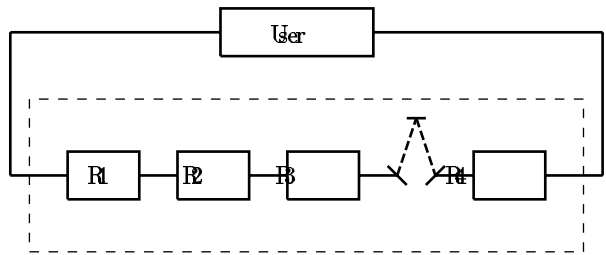

Figure 4: The VLAN Configuration

### 6.1 Case I: Very Large Area Network

The first network configuration is a satellite link with several terrestrial links. Satellite networks are now called very large area networks (VLAN) and are important since most large networks generally consist of several wide area networks (WAN) and local area networks (LAN) connected together via satellite links. A queueing model of the configuration simulated is shown in Figure 4.

The queueing model of the network consists of four servers with deterministic service times of 2, 5, 3, and 4 units of time. The satellite link is represented by a fixed (regardless of window) delay of 62.5 units of time. All service times are relative to source service time which therefore has a service time of 1. For this network, the bottleneck server's service time $t_b = 5$, and $\sum t_i = 77.5$. If the total number of packets in this network is $W$, the delay $D$ is given by:

$$D = \text{Max}\{77.5, 5W\}$$

The knee of the delay curve (see Figure 5) is at $W_{knee} = 77.5/5 = 15.5$.

A plot of window as a function of time, as obtained from simulation using the the sample scheme, is shown in the Figure 6. Notice that within 16 window adjustments, the window reaches the optimal value and then oscillates between 12 and 16. Every fourth cycle, the window curve takes an up turn at 13 (rather than at 12) because we maintain window values as $real$ numbers even though the actual number of packets sent is the nearest integer.

### 6.2 Case II: Wide Area Network

The second configuration presented is that of a terrestrial wide area network. This configuration is similar to the VLAN network except that there is no satellite link. A queueing model of the configuration is shown in Figure 7. The service times of the five servers are 2, 5, 4, and 3 time units (relative to the source). The delay with W packet circulating in the network is given by:

$$D = \text{Max}\{15, 5W\}$$

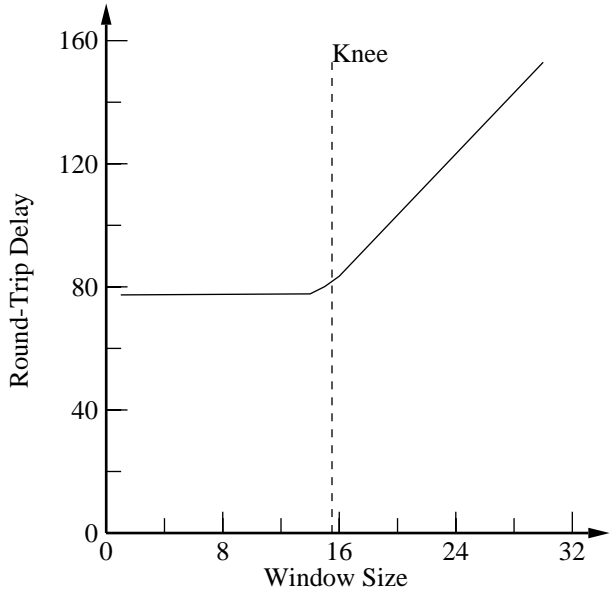

Figure 5: Round-trip delay in the VLAN Configuration.

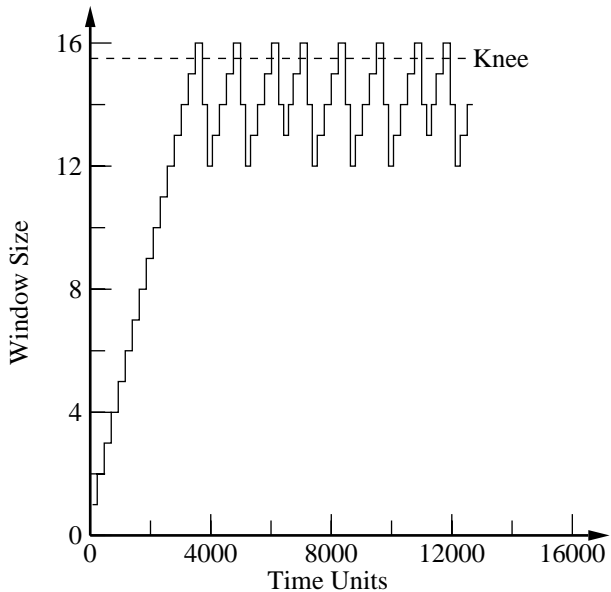

Figure 6: Window using the delay-based schema for the VLAN configuration.

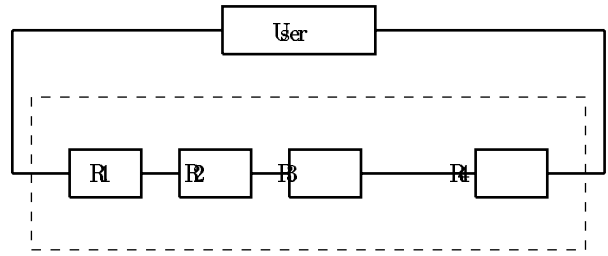

Figure 7: The WAN Configuration.

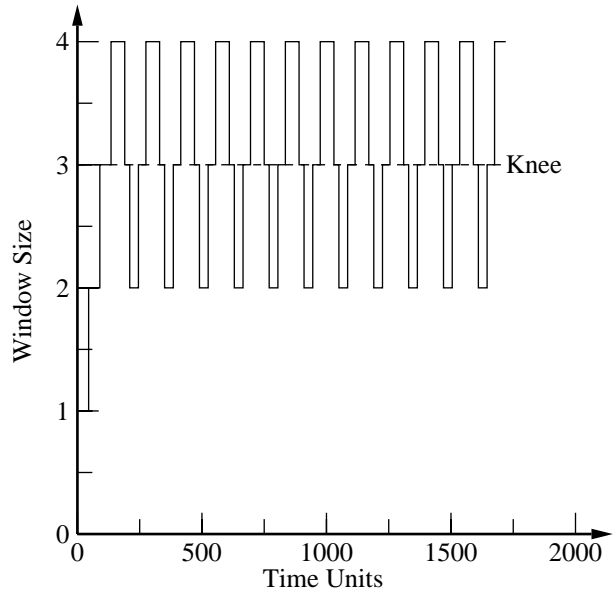

Figure 8: Window for the WAN Configuration.

The knee of the delay curve is at $W_{knee} = 3$.

Figure 8 shows the window curve as obtained using the sample schema. Once again, we see that the window oscillates closely around the knee.

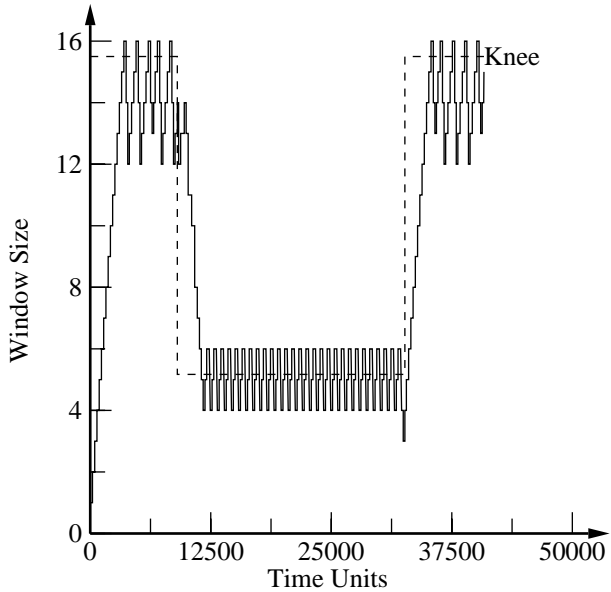

Figure 9: Responsiveness of the scheme to changes in link speeds.

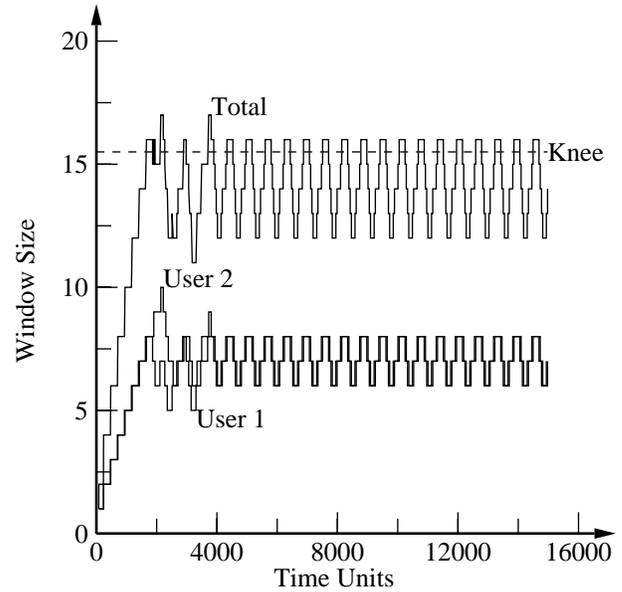

Figure 10: Performance for two users in a VAN configuration.

### 6.3 Responsiveness to Configuration Changes

Computer networks are constantly reconfiguring as links go down or come up. To test if the congestion avoidance scheme would respond to such dynamic conditions, we simulated the VAN configuration described above. We divided the input packet stream into three equal parts. During the middle part we changed the bottleneck router speed by a factor of 3 so that the optimal window size changed from 15.5 to 5.17. As seen in Figure 9, the delay based scheme did respond very well to this change. In the third part of the stream, we changed the bottleneck servers speed back to original and once again the window curve came back to the optimum.

### 6.4 Fairness

Figure 10 shows the performance for the VAN network with two users. The optimal window per user in this case is 7.75 and as seen from the figure both users have windows that oscillate between 6 and 8. The total (sum of the two) window oscillates between 12 and 16.

### 6.5 Any Initial Window

Since the scheme is responsive and adapts to changes in the network configuration, the initial window where a user starts is irrelevant. We verified this by using a VAN network with the user starting at a very high window. As shown in Figure 11, the user quickly comes down to the knee.

### 6.6 Convergence under Heavy Congestion

Figure 12 shows window curve for a highly congested WN configuration with nine users. The knee capacity of the path is only three. The optimal window per user is one-third. Since the minimum window size is 1, the users keep oscillating between 1 and 2 and total window oscillates between 9 and 18.

Many alternative decision functions were rejected as a result of divergence for this configuration. Figure 13 shows simulation results for such a diverging case with users trying to optimize their local power (rather than simply checking NG to be zero). The users discover that to optimize their local power they need windows at least as large as the sum of the other users. This leads to a case where the mean window of the users keeps going up without bound.

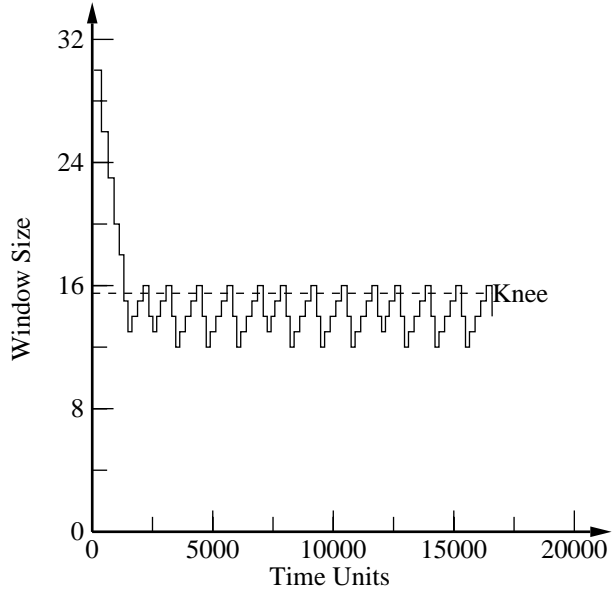

Figure 11: The window converges to the knee capacity regardless of the starting window

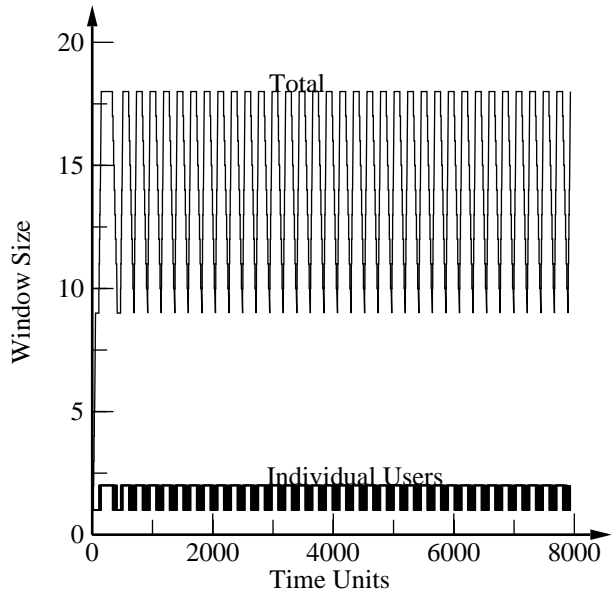

Figure 12: The schema converges for heavily congested networks.

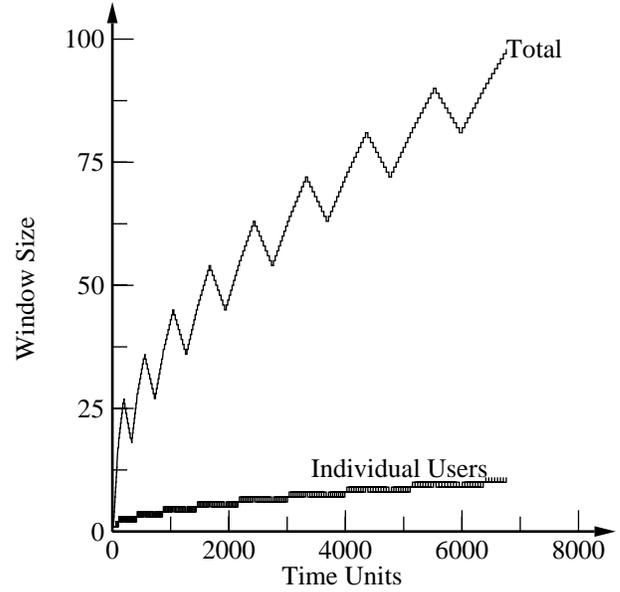

Figure 13: A decision function that leads to divergence under heavy congestion. This decision function was rejected.

## 7 FEATURES OF THE SCHEME

The design of the schema described here was based on a number of goals that we had determined beforehand. Below we show how the proposed schema meets these goals at least for deterministic networks.

1. Zero network overhead: There is no overhead on intermediate systems. This schema does not require intermediate systems to measure their loads or queue lengths. Their resources can be dedicated for packet forwarding rather than feedback.

2. No new packets: Unlike source quench schema or choke packet schema [18], this schema does not require any new packets to be injected into the network during overload or underload.

3. No change in packet headers: The schema will work in all networks with their existing packet formats.

4. Distributed control: The schema is distributed and works without any central observer.

5. Dynamism: Network configurations and traffic vary continuously. Nodes and links come

up and down and the load placed on the network by users varies widely. The optimal operating point is therefore a continuously moving target. The proposed scheme dynamically adjusts its operation to the current optimal point. The users continuously monitor the network by changing the load slightly below and slightly above the optimal point and verify the current state by observing the feedback.

6. Minimum oscillation: The increase amount of 1 and decrease factor of 0.875 have been chosen to minimize the amplitude of oscillations in the window sizes.

7. Convergence: If the network configuration and workload remain stable, the scheme brings the network to a stable operating point.

8. Low parameter sensitivity: While comparing various alternatives, we studied their sensitivity with respect to parameter values. We discarded several alternatives simply because their performance was highly sensitive to the setting of a parameter value.

9. Information entropy: Information entropy relates to the use of feedback information. We want to get the maximum information across with the minimum amount of feedback. By using implicit feedback, this scheme allows several bits worth of information to be obtained without using any physical bits.

10. Dimensionless parameters: A parameter that has dimensions (length, mass, time) is generally a function of network speed or configuration. A dimensionless parameter has wider applicability. The window update frequency, window increase amount, and window decrease factor are all dimensionless. We specifically rejected alternatives that required using parameters such as minimum delay or maximum gradient because such parameters have dimensions and would be valid only for networks of certain bandwidths and extents.

11. Configuration independence: No prior knowledge of the network configuration, number of hops, presence or absence of satellite links, etc. is required.

Most of the discussion in this paper centers around window-based flow control mechanisms. However, we must point out that this is not a requirement. The congestion avoidance algorithm and concepts can be easily modified for other forms of flow control such as rate-based flow control, in which the sources must send at a rate lower than a maximum rate (in packets/second or bytes/second) specified by the destination. In this case, the users would adjust rates based on the delay experienced.

In developing the scheme proposed here, we assumed that round-trip delay can be estimated. This is possible only if packets are acknowledged explicitly or implicitly (by acknowledgment bits or by response to a request). Not every packet needs to be acknowledged though. Most networking architectures, including DNA, use only one timer to measure the round-trip delay while a number of packets are outstanding. This is sufficient. The impact of withholding acknowledgment arbitrarily needs further work. But, if the delay introduced is fixed (regardless of the window), the effect is similar to that of a satellite link, and the scheme is expected to work.

# 8 Areas For Further Research

The main purpose of this paper is to introduce researchers in this area to the possibility of designing delay-based schemes for congestion avoidance. The ideas presented here are only a beginning. Much remains to be done to make it a practical scheme. Some of the areas needing further research are:

1. Alternative decision functions
2. Additional information
3. Extension to probabilistic networks
4. Alternative optimality criteria

In this section, we explain the above areas and describe possible solution approaches briefly. However, all statements in this section are speculative, and some may eventually turn out to be false.

## 8.1 Alternative Decision Functions

We used NG as the decision function. Other possibilities are:

1. Intercept: Given delays at two different window values, one can fit a straight line of the form

$$D = aW + b$$

Here, $a$ is the gradient and $b$ is the intercept of the line. Before the knee, the intercept is close to the delay $D$, while after the knee, the intercept is close to zero.

2. Intercept/Gradient Ratio: Ratio $b/a$ is large before the knee but very small after the knee.

3. Delay at Minimum Window: Before the knee, the delay is close to the delay at $W=1$, while after the knee, it several times the delay at $W=1$. In networks that can modeled as a closed queueing network of several M/M/1 servers, the delay at the knee is approximately twice the delay without any queueing. Thus, if we measure the delay at $W=1$, we can continue increasing the window till the delay is twice this amount.

It should be obvious that several other combinations of NU, intercept, gradient, and minimum delay can also be used.

## 8.2 Additional Information

In developing the schema proposed in this paper, we followed a *pure* black-box approach by assuming no knowledge whatsoever about the path. Additional information is sometimes available and can be useful. Examples of such information are:

1. Number of users sharing the path: If the number of users $n$ sharing the path is known, it is possible to reach close to social optimum using local power. If *each* user uses only $1/(2n-1)$ of the window predicted by the selfish optimum, i.e.,

$$W_i \leftarrow \frac{1}{2n-1}\left(\frac{\alpha}{1+\alpha}\right)\left(\frac{D}{\frac{\partial D}{\partial W_i}}\right)$$

then, it can be shown that starting from any initial condition the window will eventually converge to a fair and socially optimal value so that

$$W_i = W_j = \frac{1}{n}\left(\frac{\alpha}{1+\alpha}\right)\left(\frac{D}{\frac{\partial D}{\partial W_i}}\right) \forall i,j$$

It is possible to statically select $n$ or make it a network parameter set by the network manager. In this case, the performance is slightly suboptimum during periods when actual number of users is below $n$, and the schema may diverge during periods when the number of users exceeds $n$. The divergence can be controlled by setting a limit $W_{max}$.

2. Minimum delay: If minimum delay (delay through a path with no queueing anywhere) is known, we can estimate the current load of other users on the network from current delay and thereby try to achieve the social optimum. The gradient of the delay-window curve, if nonzero, is proportional to the bottleneck service time, and the minimum delay is equal to the sum of all service times. These two allow us to compute the knee capacity of the path. The difference in delay at $W_i=1$ and minimum delay is proportional to the load put by other users on the network. A user can thus compute its share of the load to achieve social optimum.

Many networking architectures assign *cost* to network links based on their speed and use it to select the optimal path. In networks with very fast links, the service times at the switching nodes determine the optimality of a path and not the link speed. Thus, if cost were assigned to all servers (links as well as switches) based on their packet service time, the cost of a path would be a measure of the minimum delay.

## 8.3 Extension to Probabilistic Networks

The key area for further research is to extend the schema for probabilistic networks in which the service time per packet at each server is a random variable. Without that extension, the schema is not yet ready for practical implementations.

If we allow the service times of the servers to be random variables with a probability distribution, the round-trip delay becomes random too. Any decision based on the delay then has a certain probability of being wrong. There are several alternatives to handle this problem.

1. Signal Filtering: A straightforward extension of the schema to random service times would be to take several samples of delay at a given window and estimate the mean and confidence interval of NU.

One problem with straight filtering is that delay is not a random *variable*, it is a random *process*. A random variable is characterized by a probability distribution function with parameters that do not change with time. A random process is characterized by a probability distribution function whose parameters change with time. These changes are caused by changes in network configuration or load. Unless a stochastic process is *stationary*, the time average (average of samples taken at different times) is not identical to space average (average of several samples taken at the same time). In any case, all averaging should be such that the recent samples have more impact on the decision making than the old samples. An exponentially-weighted averaging is therefore preferable to a straightforward summation of all samples taken for the same window

2. Decision Filtering: Another approach to handle randomness is to make several, say, $2k+1$ decisions each based on a single sample. All decisions will not be identical. Some will ask the user to increase while the others will ask it to decrease the window. The final action taken will be as dictated by the majority. The probability of errors can be minimized by increasing $k$. Let $p$ be the probability of correct decision based on one sample. Then, probability of correct decision based on $2k+1$ samples would be:

$$\sum_{i=k+1}^{2k+1} \binom{2k+1}{i} p^i (1-p)^{2k+1-i}$$

Similarly, the probability of incorrect decision is:

$$\sum_{i=0}^{k} \binom{2k+1}{i} p^i (1-p)^{2k+1-i}$$

Again, the decisions may be 'aged-out' and recent decisions may be given a higher weight than earlier ones.

3. Sequential Testing: In the deterministic version of the delay scheme, we check to see if NDG is zero. In the probabilistic version, we would need to change this to a statistical hypothesis test with a specified confidence level. We may design a sequential testing procedure such that after $k$ samples, the test asks us to increase, decrease, or to take one more sample.

4. Goal Change: For deterministic cases, NDG of delay-window curve is zero to the left of the knee. This is not always true for probabilistic cases. For example, for a *balanced* network of $h+1$ identical M/M/1 servers in a cycle, the average round-trip delay with $\sum_{i=1}^{n} W_i$ packets circulating in the cycle is:

$$D = \left( h + \sum_{i=1}^{n} W_i \right) t_b$$

where $t_b$ is the service time of each server. For this case, the delay curve is a single straight line, and there is no visible knee on the curve. Mathematically though, the knee can be determined as follows. The system power is:

$$P = \frac{T^\alpha}{D} = \frac{(\sum_{i=1}^{n} W_i)^\alpha}{D^{1+\alpha}}$$
$$= \frac{(\sum_{i=1}^{n} W_i)^\alpha}{\{(h + \sum_{i=1}^{n} W_i) t_b\}^{1+\alpha}}$$

It is maximum at:

$$\sum_{i=1}^{n} W_i = h$$

The following holds at the optimal point:

$$D = 2ht_b = 2D_0$$

Here $D_0$ is the average minimum delay on the network with no packets circulating. Thus, the ratio of the delay to minimum delay rather than NDG is a better indicator of the knee for such a case.

The exponential distribution of service time assumed in the above analysis is only for analytical convenience. In most practical networks, the service times have a variance much smaller than that implied by the exponential distribution. In the past, one reason for variability of service time used to be the byte-by-byte handling of packets such that the service time was proportional to the packet length. Current trend is to get away from such handling, and the packet service times are getting closer to the constant distribution and away from the exponential.

## 8.4 Alternative Optimality Criteria

The difficulty in finding a distributed scheme for social optimum is partly due to the definition of the 'optimum' using power. Jaffe [8] has shown that the network power is nondecentralizable. This, in fact, has been the strongest argument against use of power as a goal, and it has lead researchers to look for other functions which can be decentralized. For example, the *new power* function proposed by Selga [21] achieves its maximum when the delay is a multiple (say, twice) the minimum delay. This requires knowing minimum delay of the path. However, if the minimum delay is known then we may be able to extend the delay based approach as discussed earlier in this section.

## 8.5 Game Theory

The social vs selfish conflict suggests that game theory may be able to help us in changing the optimization problem from a competitive game to a cooperative game. Most cooperative games (or team efforts) require considerable exchange of information. Sanders [22], for example, proposes using an incentive scheme to prevent the users from getting into a selfish mode. However, her resource allocation mechanism uses a central node to collect information about network state. A distributed version of the mechanism would entail considerable overhead.

## 9 Summary

Round-trip delays through the network are an implicit indicator of load on the network. Using these provides a way for congestion avoidance in heterogeneous networks. Even in homogeneous networks, this solves the problem of congestion at resources, such as bridges, which do not operate at the architectural layer at which explicit congestion feedback can be provided. Also, it has the desired property of putting zero overhead on the network itself.

We have described a sample scheme in which the sources use round-trip delay as the only feedback available to control their load on the network. The key limitation of the scheme is that it works only for deterministic networks, i.e., networks in which packet service time per packet is constant. Using a simulation model, we have tried many different deterministic configurations and scenarios. We have found the scheme to be convergent, fair, optimum, and adaptive to network configuration changes.

One of the key issues during the design of this scheme was selfish optimum versus social optimum. We rejected several alternatives that achieved selfish optimum and caused a race condition leading to divergence.

The results of our initial efforts in achieving congestion avoidance using round-trip delays are encouraging. However, much remains to be done to make it a practical scheme for implementation in real networks where the service times are random and where users are competing rather than cooperating. Extending the approach to probabilistic networks, using game theoretic concepts or by getting additional information about the network, is a promising direction for further research in this area.

## 10 Acknowledgments

Many architects and implementors of Digital's networking architecture participated in a series of meetings over the last four years where the ideas presented here were discussed and improved. In particular, we would like to thank Linda Wright for encouraging us to work in this area, to Bill Hawe, and Tony Lauck for valuable feedback, and to George Varghese for suggesting that we look into black-box approaches to congestion avoidance.

## References

[1] V Ahuja, "Routing and Flow Control in System Network Architecture," IBM System Journal, Vol. 18, No. 2, 1979, pp. 298 - 314.

[2] W Bux and D Grillo, "Flow Control in Local-Area Networks of Interconnected Token Rings," IEEE Transactions on Communications, Vol. COM-33, No. 10, October 1985, pp. 1058-66.

[3] Dah-Ming Chiu and Raj Jain, "Analysis of Increase/Decrease Algorithms For Congestion Avoidance in Computer Networks," Digital Equipment Corporation, Technical Report DEC-TR-509, August 1987, To be published in Computer Networks and ISDN Systems.

[4] Digital Equipment Corp., "DECnet Digital Network Architecture (Phase V) General Description," Order NO EK-DNAPV-GD, September 1987.

[5] A Gessler, J Haenle, A Konig and E Pade, "Free Buffer Allocation - An Investigation by Simulation," Computer Networks, Vol. 1, No. 3, July 1978, pp. 191-204.

[6] International Organization of Standardization, "ISO 8073: Information Processing Systems -

Open Systems Interconnection - Connection Oriented Transport Protocol Specification," July 1986.

[7] M. Irland, "Buffer Management in a Packet Switch," IEEE Trans. on Comm., Vol. COM-26, March 1978, pp. 328-337.

[8] J. M. Jaffe, "Flow Control Power is Nondecentralizable," IEEE Transaction on Communications, Vol. COM-29, No. 9, September 1981, pp. 1301-1306.

[9] Raj Jain, Dah-Ming Chiu, and William Hawe, "A Quantitative Measure of Fairness and Discrimination for Resource Allocation in Shared Systems," Digital Equipment Corporation, Technical Report DEC-TR-301, September 1984.

[10] Raj Jain, "Using Simulation to Design a Computer Network Congestion Control Protocol," Proc. Sixteenth Annual Pittsburgh Conference on Modeling and Simulation, Pittsburgh, PA, April 25-26, 1985, pp. 987-993.

[11] Raj Jain, "Divergence of Timeout Algorithms for Packet Retransmission," Proc. Fifth Annual International Phoenix Conf. on Computers and Communications, Scottsdale, AZ, March 26-28, 1986, pp. 174-179.

[12] Raj Jain and William Hawe, "Performance Analysis and Modeling of Digital's Networking Architecture," Digital Technical Journal, No. 3, September 1986, pp. 25-34.

[13] Raj Jain and Shawn Routhier, "Packet Trains - Measurements and a New Model for Computer Network Traffic," IEEE Journal on Selected Areas in Communications, Vol. SAC-4, No. 6, September 1986, pp. 986-995.

[14] Raj Jain, "A Timeout-Based Congestion Control Scheme for Window Flow Controlled Networks," IEEE Journal on Selected Areas in Communications, Vol. SAC-4, No. 7, October 1986, pp. 1162-1167.

[15] Raj Jain, K. K. Ramakrishnan, and Dah-Ming Chiu, "Congestion Avoidance in Computer Networks with a Connectionless Network Layer," Digital Equipment Corporation, Technical Report DEC-TR-506, August 1987.

[16] Raj Jain and K. K. Ramakrishnan, "Congestion Avoidance in Computer Networks with a Connectionless Network Layer: Concepts, Goals and Methodology," Proc. IEEE Computer Networking Symposium, Washington, D C, April 1988, pp. 134-143.

[17] L. Kleinrock, "Power and Deterministic Rules of Thumb for Probabilistic Problems in Computer Communications," in Proc. Int. Conf. Comm., June 1979, pp. 43.1.1-10.

[18] J. C. Majithia, et al, "Experiments in Congestion Control Techniques," Proc. Int. Symp. Flow Control Computer Networks, Versailles, France, February 1979.

[19] K. K. Ramakrishnan and Raj Jain, "An Explicit Binary Feedback Scheme for Congestion Avoidance in Computer Networks with a Connectionless Network Layer," Proc. ACM SIGCOMM'88, Stanford, CA, August 1988.

[20] K. K. Ramakrishnan, Dah-Ming Chiu and Raj Jain, "Congestion Avoidance in Computer Networks with a Connectionless Network Layer. Part IV A Selective Binary Feedback Scheme for General Topologies," Digital Equipment Corporation, Technical Report DEC-TR-510, August 1987.

[21] J. M. Selga, "New Flow Control Power is Decentralizable and Fair," Proc. IEEE INFOCOM'84, pp. 87-94.

[22] B. A. Sanders, "An Incentive Compatible Flow Control Algorithm for Fair Rate Allocation in Computer/Communication Networks," Proc. Sixth International Conf. on Distributed Computing Systems, 1986, pp. 314-320.